\documentclass[12pt]{iopart}
\usepackage{iopams,graphicx}  
\usepackage{bm}
\begin{document}

\title{Multiple Scattering Formulation of Two Dimensional Acoustic and Electromagnetic Metamaterials}

\author{Daniel Torrent and Jos\'e S\'anchez-Dehesa}

\address{Wave Phenomena Group, Universidad Polit\'ecnica de Valencia, C/Camino de Vera s/n, 46022 Valencia, Spain}
\ead{jsdehesa@upvnet.upv.es}
\begin{abstract}
This work presents a multiple scattering formulation of two dimensional acoustic metamaterials. It is shown that in the low frequency limit multiple
scattering allows us to define frequency-dependent effective acoustic parameters for arrays of both ordered and
disordered cylinders. This formulation can lead to both positive and negative acoustic parameters, where the acoustic parameters
are the scalar bulk modulus and the tensorial mass density and, therefore, anisotropic wave propagation is allowed with both positive or negative
refraction index. It is also shown that the surface fields on the scatterer are the main responsible of the anomalous behavior of the effective medium,
therefore complex scatterers can be used to engineer the frequency response of the effective medium, and some examples of application to different
scatterers are given. Finally, the theory is extended to electromagnetic wave propagation, where Mie resonances are found to be the responsible of the metamaterial behavior. As an application, it is shown that it is possible to obtain metamaterials with negative permeability and permittivity tensors by arrays of all-dielectric cylinders and that anisotropic cylinders can tune the frequency response of these resonances. 
\end{abstract}
\maketitle

\section{Introduction}
Fluid-like metamaterials or metafluids with negative constitutive parameters offer a new insight into acoustic wave propagation. Single negative metamaterials (SNM), in which either the mass density or the bulk modulus is negative \cite{liu2000locally,fang2006ultrasonic,yang2008membrane}, can be used, for example, designing of surface-like acoustic lens to overcome the diffraction limit \cite{ambati2007surface,deng2009theoretical} or for the design of acoustic panels \cite{yang2010acoustic}. Double negative metamaterials (DNM)\cite{li2004double, cheng2008one,lee2010composite} present negative refraction \cite{veselago1968electrodynamics,pendry2000negative,smith2004metamaterials,shalaev2007optical} and, as it is well known from electromagnetic wave theory, they can also be used to increase the resolution of conventional lens \cite{jacob2006hyperlens,zhang2008superlenses,li2009hyperlens}. In general, anisotropic fluid-like metamaterials with acoustic parameters both positive and negative are necessary in the field of transformation acoustics for the design of several types of acoustic devices based on sound propagation \cite{cummer2007one,cummer2008material,chen2010acoustic,bin2010noise,yang2010super}.
\par 
The existence of frequency ranges in which the effective medium presents negative constitutive parameters is related with the subwavelength resonances of the individual scatterers that constitute the metamaterial, being these due to soft-scatterer resonances \cite{li2004double,liu2005analytic,ding2007metamaterial} or due to Helmholtz-like resonances \cite{hu2005two,wang2008acoustic,hu2008homogenization, cheng2008one,fang2006ultrasonic}. The same phenomenon
is found in electromagnetic waves under the name of Mie resonances \cite{peng2007experimental,schuller2007dielectric,vynck2009all,chern2010effective}, and they present an alternate way of design electromagnetic 
metamaterials to that offered by split ring resonators \cite{pendry1999magnetism} or metallodielectric composites \cite{shalaev2007optical}, which have been the dominant structures so far. Therefore, 
metamaterials based on the resonances of the individual scatterers are important not only for acoustic but also for electromagnetic metamaterials.
\par 
It is known that a monopolar resonance in the scatterer is the responsible of the negative bulk modulus, and that a dipolar one is the
responsible of the negative mass density \cite{li2004double}. However, the full effect of the ensemble of scatterers that constitute the effective medium has been partially explained only, as multiple scattering effects or anisotropic lattices have not been considered yet.
\par 
In this work, we give a comprehensive description of multiple of acoustic metamaterials by using a multiple scattering formulation. It is based on our previous results on homogenization of sonic crystals \cite{torrent2006homogenization,torrent2006effective,torrent2008anisotropic}. This formulation describes, in the long wavelength limit, an ensemble of orderer or disordered scatterers as an effective medium with frequency-dependent acoustic parameters, which are shown to be negative in certain frequency regions. The frequency-dependent parameters are given in terms of both the lattice symmetry and the surface fields on the scatterers, showing that these fields are the quantities that we have to manage in order to engineer the frequency response of our effective medium. 
\par 
Therefore this formulation allow us, from one side, summarize all the previous
results regarding SNM and DNM, from the other side, extend the theory to any type of radially symmetric  scatterer and to
non symmetric lattices, like rectangular arrangements in place of square or hexagonal ones, which are the more usually studied. Also, we show that
the regions in which the metamaterial presents divergent or negative acoustic parameters is mainly a function of the surface fields of the scatterers, opening therefore the possibility of engineer the scatterer in order to increase the frequency region in which the anomalous behavior occurs.
\par 
The paper is organized as follows: In section \ref{sec:resonances} the concept of quasi-static resonance is introduced, showing how these resonances lead to scatterers
with locally negative parameters. Section \ref{sec:scatterers} analyzes several examples of scatterers with locally negative parameters, showing that complex scatterers
like fluid-like shells or anisotropic fluid-like cylinders can tune the metamaterial behavior of the scatterer. After that, in Section \ref{sec:mst} the multiple scattering formulation is presented, analyzing first the case in which the scatterers have small radius (low filling fraction) and, later, the more general case, showing how anisotropy appears. In this section the theory is extended to include multipolar effects, but we see that they are not very important in principle. Finally, the application
of the theory to electromagnetic waves is explained in section \ref{sec:oem}. The paper ends with a summary section.

\section{Quasi-Static Resonances and Locally Negative Parameters}
\label{sec:resonances}
Homogenization theories for assembles of scatterers are based on the low wavenumber (long wavelength) expansions of the fields in both the background and the 
scatterers. When working with metamaterials we assume that the wavenumber in the background is asymptotically small though we let the wavenumber inside the scatterer still be finite. Physically it means that outside the scatterers the wave field propagates through an
effective medium but it is still allowed to the scatterers to have complex internal scattering processes, which will lead to locally (i.e., in a narrow frequency region) negative parameters, as will be explained in the following sections.
\par
The simpler example of these scatterers, and the first studied in the next section, is the homogeneous fluid-like scatterer. If the speed of sound
inside this scatterer is much smaller than that of the background, $c_{a}<<c_{b}$, we will have that, for a given frequency $\omega$, the wavelength
inside the scatterer $\lambda_{a}$ will also be much smaller than that in the background $\lambda_{a}<<\lambda_{b}$. Thus, outside the scatterer
the field will be a function of $k_{b}=\omega/c_{b}$, which is a slow oscillating function, while inside the scatterer the fields will be a function of $k_{a}=\omega/c_{a}$ and, therefore, it is a rapidly oscillating function. As we are in the low frequency limit we expect the medium behave as a homogeneous effective
medium with constant parameters, but, due to the fields inside the scatterer, we will find that our effective medium has frequency-dependent parameters.
\par
Next section analyzes this effect rigorously and for several type of scatterers, showing how complex scatterers present a frequency behavior that, as will be seen in Section \ref{sec:mst}, can lead to metamaterials with negative constitutive parameters.
\section{Acoustic Scatterers with Locally Negative Parameters}
\label{sec:scatterers}
The wave equation for the pressure field in an inhomogeneous fluid is given by \cite{morse1986theoretical}
\begin{equation}
\label{eq:ac_wave_equation}
\nabla (\rho^{-1}(\bm{r})\nabla P(\bm{r}))+\frac{\omega^2}{B(\bm{r})}P(\bm{r})=0
\end{equation}
where $\rho(\bm{r})$ and $B(\bm{r})$ are the mass density and bulk modulus, respectively, of the fluid material. The problem considered here is reduced to points in the 
$x-y$ plane, that in polar coordinates are $\bm{r}=(r,\theta)$. We also assume that a radially symmetric scatterer of radius $R_{a}$ with some inhomogeneous parameters $\rho(r)$ and $B(r)$ is embedded into a fluid background with acoustic parameters $\rho_{b}$ and $B_{b}$.
\par
This is a canonical scattering problem whose solution outside the scatterer is given in terms of Bessel and Hankel functions \cite{morse1986theoretical}, 
\begin{equation}
P(r,\theta;\omega)=\sum_{q}A_{q}^{0}\left[J_{q}(k_{b}r)+T_{q}H_{q}(k_{b}r)\right]e^{iq\theta}\quad,\quad r>R_{a}
\end{equation}
with $k_{b}^{2}=\omega^{2}\rho_{b}/B_{b}$. The coefficients $A_{q}^{0}$ are determined by the incident field, and the response of the scatterer is described by the matrix elements $T_{q}$. This matrix
is obtained by solving the wave equation \eref{eq:ac_wave_equation} inside the scatterer and applying boundary conditions at $r=R_{a}$, which
are the continuity of the pressure field and that of the normal component of the particle velocity,
\numparts
\begin{eqnarray}
P(R_{a}^{+})&=&P(R_{a}^{-})\\
\frac{1}{\rho_{b}}\partial_{r}P(R_{a}^{+})&=&\frac{1}{\rho(R_{a})}\partial_{r}P(R_{a}^{-}).
\end{eqnarray}
\endnumparts
\par
Since the scatterer is radially symmetric, and the parameters $\rho$ and $B$ depend only on the radial coordinate, the field
inside the scatterer is expressed in a Fourier series of the form
\begin{equation}
P(r,\theta;\omega)=\sum_{q}B_{q}(\omega)\psi_{q}(r;\omega)e^{iq\theta},
\end{equation}
where the eigenfunctions $\psi_{q}(r;\omega)$ are solutions of the radial part of \eref{eq:ac_wave_equation} in cylindrical coordinates,
\begin{equation}
\label{eq:wave_radial}
\frac{\rho(r)}{r}\partial_{r}\left(\frac{r}{\rho(r)}\partial_{r}\psi_{q}(r;\omega)\right)+\left(\omega^2\frac{\rho(r)}{B(r)}-\frac{q^2}{r^{2}}\right)\psi_{q}(r;\omega)=0.
\end{equation}
From this equation, after applying the boundary conditions, the general form for the T matrix is obtained
\begin{equation}
T_{q}=-\frac{\chi_{q}J'_{q}(k_{b}R_{a})-J_{q}(k_{b}R_{a})}{\chi_{q}H'_{q}(k_{b}R_{a})-H_{q}(k_{b}R_{a})}\quad,\quad \chi_{q}=\frac{\rho(R_{a})}{\rho_{b}}\frac{\psi_{q}(R_{a};\omega)}{\partial_{r}\psi_{q}(R_{a};\omega)}k_{b}
\end{equation}
\par
This $T$ matrix allows to distinguish the contribution of the background from that of the scatterer. The background contribution is described by the Bessel and Hankel functions, while the scatterer contribution is described by the function $\chi_{q}$. Standard multiple scattering homogenization theory is based on the asymptotic forms of all these functions to derive, by means of the monopolar and dipolar terms ($q=0$ and $q=1$), the effective medium properties, as explained in \cite{torrent2006homogenization,torrent2006effective}. However, metamaterial behavior can appear, as it is demonstrated
below, in the regime where only Bessel and Hankel functions of the background take their asymptotic forms.
\par
Let us then consider that the arguments of Bessel and Hankel functions are small ($k_{b}R_{a}<<1$), and use their asymptotic forms \cite{abramowitz1964handbook}. The monopolar component of the T matrix is
\begin{equation}
\label{eq:lowT0}
T_0\approx \frac{i\pi R_a^2k_b^2}{4}\frac{1+\frac{1}{2}k_bR_a\chi_0}{\frac{k_b^2R_a^2}{2}\ln k_bR_a-\frac{1}{2}k_bR_a\chi_0},
\end{equation}
where the logarithmic term in the denominator is obviously equal to zero in the low frequency limit. However it cannot be neglected
when dealing with metamaterials. This term has been omitted in most of preceding works about acoustic and electromagnetic metamaterials, but
it contributes considerably to their effective parameters.
\par
Equivalently, the dipolar component of the T matrix is
\begin{equation}
\label{eq:lowT1}
T_1\approx\frac{i\pi R_a^2}{4}\frac{\chi_1/k_bR_a-1}{\chi_1/k_bR_a+1}k_b^2.
\end{equation}
Since we expect this scatterer behaves as a homogeneous scatterer with acoustic parameters $\rho_{a}$ and
$B_{{a}}$, the matrix elements should have the form
\numparts
\begin{eqnarray}
T_0\approx \frac{i\pi R_a^2k_b^2}{4}\left[\frac{B_{b}}{B_{a}}-1\right]\\
T_1\approx\frac{i\pi R_a^2}{4}\frac{\rho_{a}-\rho_{b}}{\rho_{a}+\rho_{b}}k_b^2.
\end{eqnarray}
\endnumparts
The comparison of these expressions with those given by equations \eref{eq:lowT0} and \eref{eq:lowT1} , permits to identify the frequency-dependent bulk modulus and density as
\numparts
\begin{eqnarray}
\label{eq:rhoabaomega1}
B_{a}(\omega)/B_{b}=\frac{k_b^2R_a^2}{2}\ln k_bR_a-\frac{1}{2}k_bR_a\chi_0\\
\label{eq:rhoabaomega2}
\rho_{a}(\omega)/\rho_{b}=\chi_1/k_bR_a
\end{eqnarray}
\endnumparts
\par
Note that $B_{a}(\omega)$ and $\rho_{a}(\omega)$ are obtained from the fact that, after a scattering process (in
the long wavelength limit) we expect to extract the parameters of a homogeneous fluid-like cylinder. These parameters are functions mass density at the surface of the scatterer, $\rho(r=R_{a})$, but they also depend on the field and its derivative at the surface, that is, of $\psi_{q}(r=R_{a})$ and $\partial_{r}\psi_{q}(r=R_{a})$, respectively. These quantities are frequency-dependent and, consequently, they are the responsible of the 
frequency-dependence of parameters $B_{a}(\omega)$ and $\rho_{a}(\omega)$.
\par 
In \ref{sec:scatterers} we study three types of scatterers giving negative parameters at very narrow frequency regions. The first one is a homogenous fluid like cylinder such that $c_a<<c_b$, 
condition that grants $k_b<<k_a$. The second is also a homogeneous fluid-like cylinder but now with cylindrical anisotropy. These type of cylinders have already bee studied for cloaking devices, radial wave crystals \cite{torrent2009radial} and hyperlenses \cite{li2009hyperlens} and here we will see how they can be used to tune the resonance
of the dynamical mass density of a metamaterial. Finally the third example shows that fluid-like shells can work as Helmholtz resonators,obtaining from them negative bulk modulus, but also they can work as metamaterials with negative mass density. 
\par 
However, let us point out that we report here only the simpler examples of scatterers to realize acoustic metamaterials. Obviously more complex scatterers could be used to improve the frequency
response, but a full analysis of this type is beyond the scope of the present work.

\section{Multiple Scattering of Acoustic Waves in the Quasi-Static Limit}
\label{sec:mst}
If we have a cluster of cylindrical scatterers defined by some $\rho_{a}$ and $B_{a}$, we expect that,
in the low frequency limit (that is, for wavelengths larger than the typical scatterer distance), they behave as a homogeneous medium with effective parameters $\rho_{eff}$ and $B_{eff}$. 
\par 
The comparison among the scattering properties of the cluster and the effective scatterer is used to obtain the effective parameters for the case oflow filling fractions  
\cite{berryman1980long}. In \cite{torrent2006homogenization,torrent2006effective,torrent2008anisotropic} such scattering formulation was generalized and all the multiple scattering interactions between the scatterers were included, dealing to more general expressions which also include the possibility of having
anisotropy in the mass density of the effective medium.
\par
In the following subsections the results in \cite{torrent2006homogenization,torrent2006effective,torrent2008anisotropic} are generalized to the case of metamaterials with
frequency-dependent parameters. It is assumed that
we can replace $\rho_{a}$ and $B_{a}$ by their corresponding frequency dependent values $\rho_{a}(\omega)$ and $B_{a}(\omega)$. It will be shown
that this method is self-consistent and, therefore, it defines the correct way to explain acoustic metamaterials.
\subsection{Multipolar Interactions: The $\Delta$ Factor}
Let us consider that a cluster of scatterers are periodically arranged in the space. In the low frequency limit such a cluster behaves like an effective
fluid-like medium with effective parameters given by \cite{berryman1980long}
\numparts
\begin{eqnarray}
\label{eq:eff_B}
\frac{1}{B_{eff}(\omega)}&=&\frac{1-f}{B_b}+\frac{f}{B_a(\omega)}\\
\label{eq:eff_rho}
\rho_{eff}(\omega)&=&\frac{\rho_a(\omega)(1+f)+\rho_b(1-f)}{\rho_a(\omega)(1-f)+\rho_b(1+f)}\rho_b.
\end{eqnarray}
\endnumparts
In the above expressions the dependence on the frequency has been added under the assumption that we can use for a scatterer the frequency-dependent parameters defined by equations \eref{eq:rhoabaomega1} and \eref{eq:rhoabaomega2}.\par
While equation \eref{eq:eff_B} is valid for all filling fractions, equation \eref{eq:eff_rho} is valid for diluted clusters only (i.e. low filling fractions). In \cite{torrent2006effective} and \cite{torrent2008anisotropic} the expressions for the effective density were 
generalized for the case of high filling fractions, and all the multiple scattering terms were introduced in Equation \eref{eq:eff_rho} by means of the so
called $\Delta$ factor, leading to
\begin{equation}
\rho_{eff}(\omega)=\frac{\rho_a(\omega)(\Delta+f)+\rho_b(\Delta-f)}{\rho_a(\omega)(\Delta-f)+\rho_b(\Delta+f)}\rho_b,
\end{equation}
\par
\begin{figure}
\centering
\includegraphics{ 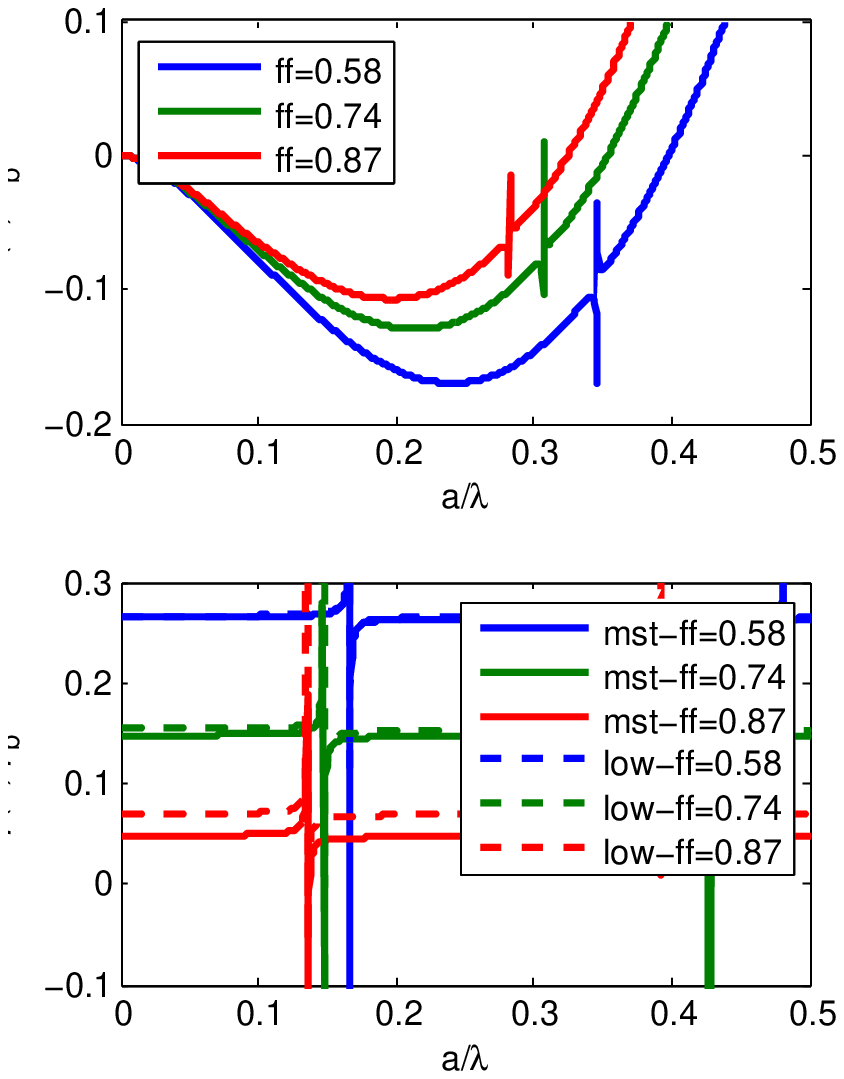}
\caption{\label{fig:Brhoair} Frequency-dependent bulk modulus for a medium composed of air cylinders in a water background.}
\end{figure}
\begin{figure}
\centering
\includegraphics{ 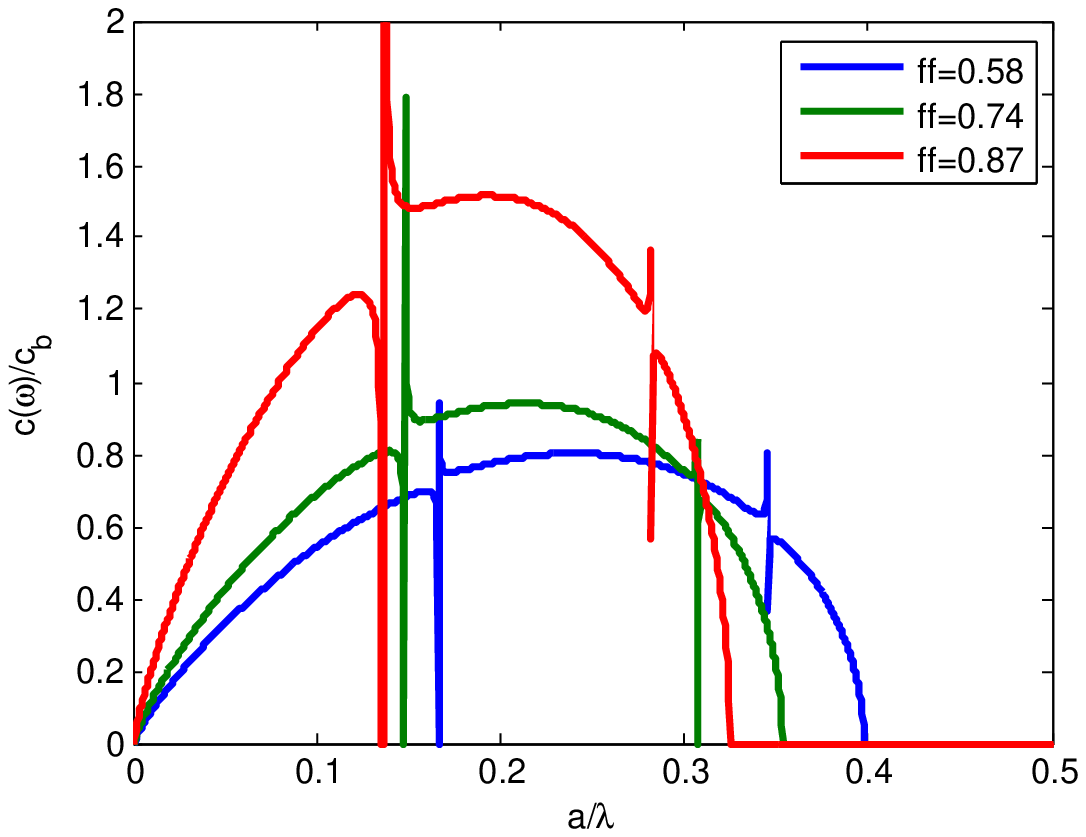}
\caption{\label{fig:cair} Imaginary component of the effective speed of sound in a medium composed of air cylinders in a water background.}
\end{figure}
The $\Delta$ factor represents a multipolar correction to the effective density expression, and it includes all the multiple scattering terms between all
the cylinders in a cluster or in an infinite lattice. Such a factor contains also information about the density of the cylinders forming the cluster, therefore
if we want to define a frequency dependent $\Delta=\Delta(\omega)$ factor and generalize the frequency dependent parameters
to all filling fractions, the frequency-dependent mass density must also be included there. However this inclusion must be made carefully.
\par

However this factor is only important for very high filling fractions and also for very strong scatterers \cite{torrent2006homogenization,torrent2006effective}, 
so we consider that such correction out of the scope of the present work; it only adds complexity to the calculation of effective parameters.
\subsection{Anisotropic Metamaterials}
Expressions in previous section have not taken into account the possibility of having anisotropy; that is, they did not consider the lattice symmetry in which cylinders are ordered in the cluster. In \cite{torrent2008anisotropic} expressions for the anisotropic mass density where found in terms of the cylinder's parameters and the lattice geometry. It can
be shown that, even neglecting the multiple scattering terms, we still have anisotropy for non-symmetric lattices, having the following expressions for the 
effective mass density tensor (the subindex ``eff'' has been omitted for clarity)
\numparts
\begin{eqnarray}
\label{eq:rhoxx}
\rho_{xx}^{-1}(\omega)=\frac{1-f^2\eta^2(\omega)(A+1)^2}{1+2f\eta(\omega)+f^2\eta^2(\omega)(1-A^2)},\\
\label{eq:rhoyy}
\rho_{yy}^{-1}(\omega)=\frac{1-f^2\eta^2(\omega)(A-1)^2}{1+2f\eta(\omega)+f^2\eta^2(\omega)(1-A^2)},
\end{eqnarray}
\endnumparts
where $A$ is the anisotropy factor explained in Appendix and $\eta(\omega)$ is defined as
\begin{equation}
\eta(\omega)=\frac{\rho_{a}(\omega)-\rho_{b}}{\rho_{a}(\omega)+\rho_{b}}
\end{equation}
It has been assumed that the lattice is oriented along the main axis and, therefore, the tensor has been previously diagonalized.
\par
The expression for the effective bulk modulus remains the same as in the previous section, so that we can obtain the effective sound speed
tensor as \cite{torrent2008anisotropic}
\begin{equation}
c^{2}_{ij}=\rho_{ij}^{-1}B_{eff}
\end{equation}
\par
\begin{figure}
\centering
\includegraphics{ 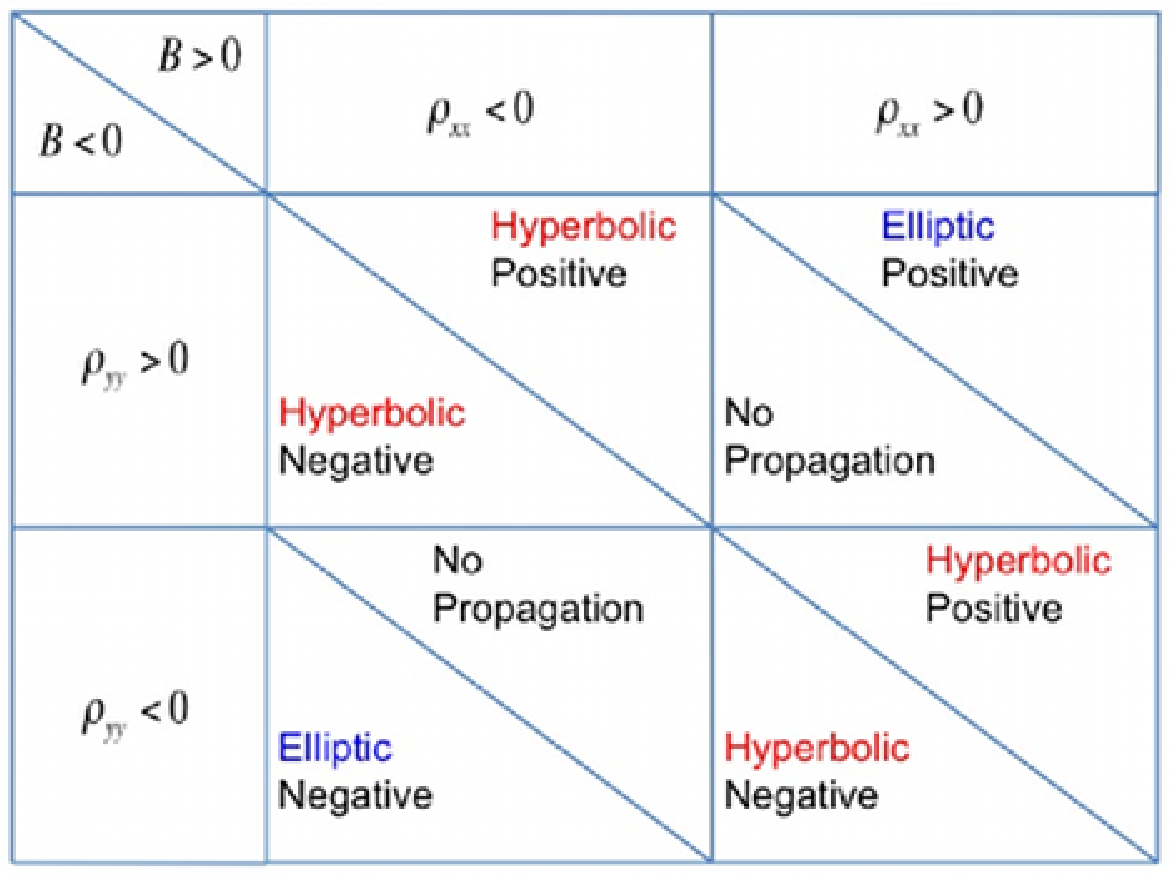}
\caption{\label{fig:table_anisotropy} Table summarizing the different propagation behavior of acoustic metamaterials.}
\end{figure}
\begin{figure}
\centering
\includegraphics{ 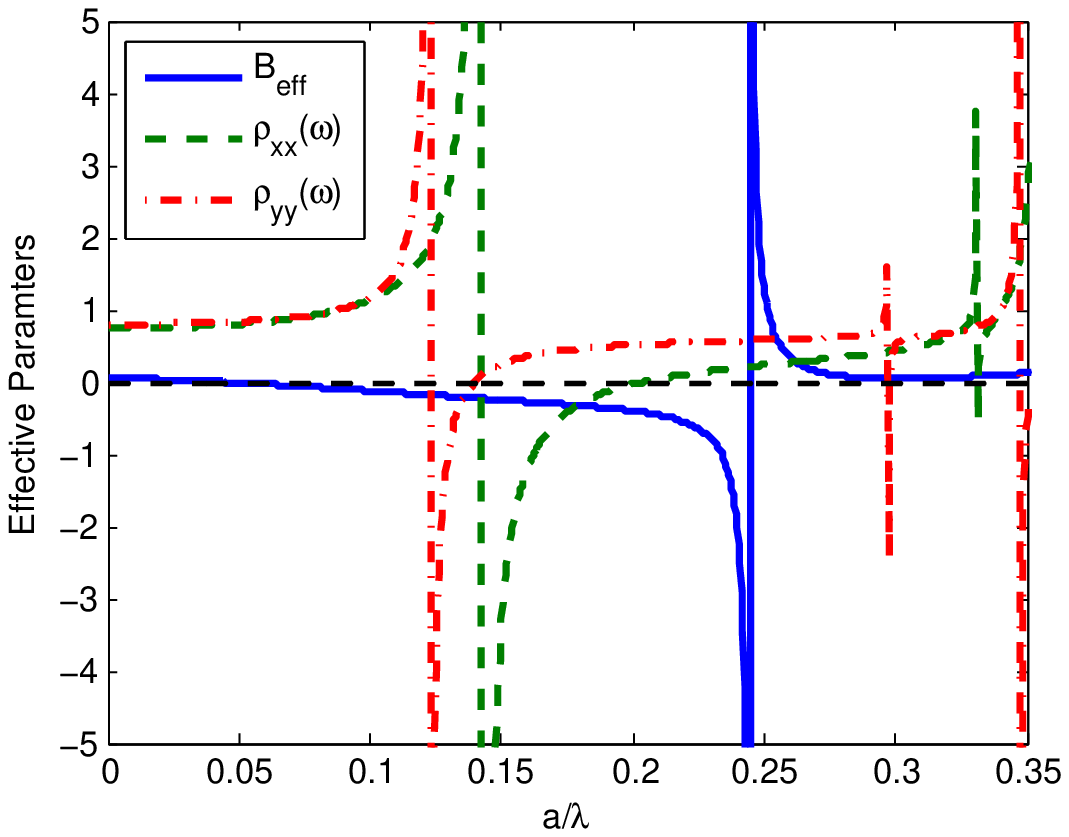}
\caption{\label{fig:rhoxxrhoyyB} Effective parameters, normalized to those of the background, as a function of frequency of a rectangular lattice of fluid-like cylinders with $b=2a$. The density, 
bulk modulus and radius of the cylinders are $\rho_{a}=0.5\rho_{b},B_{a}=0.02B_{b}$ and $R_{a}=0.49a$, respectively. We see two narrow multipolar resonances between $a/\lambda=0.3$ and 0.35, however, in that limit the homogenization hypothesis is not good. The thin horizontal dotted line is a guide for the eye.}
\end{figure}
\begin{figure}
\centering
\includegraphics{ 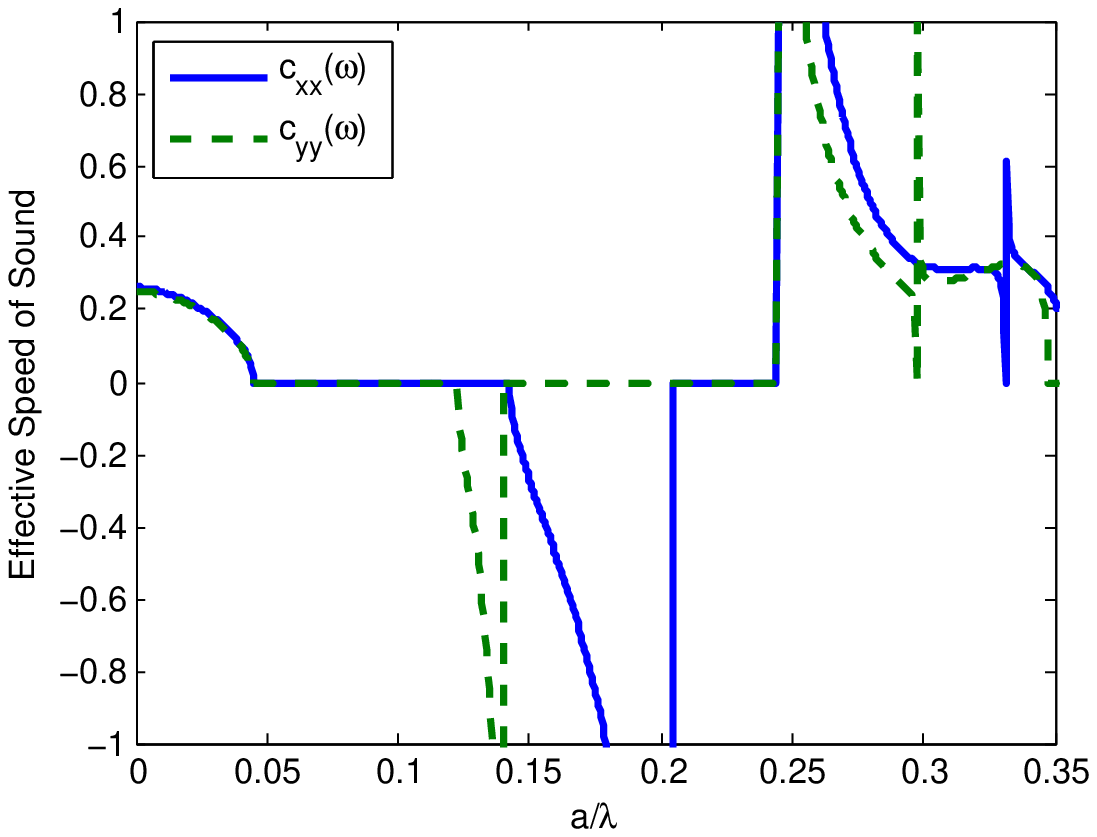}
\caption{\label{fig:cxxcyy} Real part of the effective speed of sound tensor of the medium described in Fig. \ref{fig:rhoxxrhoyyB} as a function of frequency. Note that we 
can have both components positive, or one negative and the other one imaginary (real part equal to zero in this plot). It is impossible have both components negative or one
negative and the other one positive (see text). The horizontal dotted line is a guide for the eye.}
\end{figure}
Note that the anisotropic mass density tensor can have both components with the same sign (negative or positive) or each component with a different
sign. However, as the refractive index is the square root off those terms, if we want a negative refractive index we need a negative bulk modulus too, therefore we can have only or both components of the refractive index negative, or one negative and the other one imaginary (no wave propagation). In
other words, we can have both an elliptical dispersion relation or a hyperbolic one. In Figure \ref{fig:rhoxxrhoyyB} the anisotropic mass density and the scalar bulk modulus have been 
plotted, and in Figure \ref{fig:cxxcyy} the corresponding components of the sound speed tensor are shown.

\section{Application to Electromagnetic Metamaterials}
\label{sec:oem}
The vectorial nature of electromagnetic (EM) waves makes the problem more complex, but in 2D the EM field can be decomposed into TE and TM modes, leading to the same wave equation as for scalar acoustic waves. Now $P$ in equation \eref{eq:ac_wave_equation} and \eref{eq:wave_radial} is the $z$ component of the electric (magnetic) field for TE (TM) modes, and $(\rho,B)=(\mu,\varepsilon^{-1})$  for TE modes and $(\rho,B)=(\varepsilon,\mu^{-1})$  for TM modes.
\par
Although both problems are mathematically equivalent, physically they are quite different. The numerical values and ranges of the material parameters $\rho,B$ and $\mu,\varepsilon$ are not the same in both fields, therefore it is worth to study them apart each other.
\par
Thus, for TE modes
\numparts
\begin{eqnarray}
\frac{\varepsilon_{b}}{\varepsilon_{a}^{TE}(\omega)}=\frac{k_b^2R_a^2}{2}\ln k_bR_a+\frac{k_{a}R_{a}}{2}\frac{J_{0}(k_{a}R_{a})}{J_{1}(k_{a}R_{a})}\frac{\varepsilon_{b}}{\varepsilon_{a}}\\
\nonumber\\
\frac{\mu_{a}^{TE}(\omega)}{\mu_{b}}=\frac{1}{k_{a}R_{a}}\frac{J_{1}(k_{a}R_{a})}{J'_{1}(k_{a}R_{a})}\frac{\mu_{a}}{\mu_{b}}
\end{eqnarray}
\endnumparts
For TM modes,
\numparts
\begin{eqnarray}
\frac{\mu_{b}}{\mu_{a}^{TM}(\omega)}=\frac{k_b^2R_a^2}{2}\ln k_bR_a+\frac{k_{a}R_{a}}{2}\frac{J_{0}(k_{a}R_{a})}{J_{1}(k_{a}R_{a})}\frac{\mu_{b}}{\mu_{a}}\\
\nonumber\\
\frac{\varepsilon_{a}^{TM}(\omega)}{\varepsilon_{b}}=\frac{1}{k_{a}R_{a}}\frac{J_{1}(k_{a}R_{a})}{J'_{1}(k_{a}R_{a})}\frac{\varepsilon_{a}}{\varepsilon_{b}}
\end{eqnarray}
\endnumparts
\par
These equations show that, even when the cylinders are non-magnetic ($\mu_{a}=\mu_{b}=\mu_{0}$), we can have a magnetic
response for a given frequency range. This is a well known phenomenon called mesomagnetism\cite{o2002photonic,felbacq2005theory}.
\par
Also these equations show that, as a function of frequency, the same cylinders behave in a different way for TE or TM modes, presenting different
constitutive parameters that are equal when $\omega\to0$, that is
\begin{eqnarray}
\lim_{\omega\to0}\varepsilon_{a}^{TE}(\omega)=\varepsilon_{a}^{TM}(\omega)=\varepsilon_{a}\\
\lim_{\omega\to0}\mu_{a}^{TE}(\omega)=\mu_a^{TM}(\omega)=\mu_{a}
\end{eqnarray}
\par
The effective medium made of a cluster of these scatterers will present different constitutive parameters for each of the polarizations, so that, 
applying the results of Section \ref{sec:mst},
\numparts
\label{eq:eff_electro}
\begin{eqnarray}
\varepsilon_{eff}^{TE}(\omega)&=&(1-f)\varepsilon_{b}+(1-f)\varepsilon^{TE}_{}{b}\\
\nonumber\\
\mu_{eff}^{TE}(\omega)&=&\frac{\mu_a^{TE}(\omega)(1+f)+\mu_b(1-f)}{\mu_a^{TE}(\omega)(1-f)+\mu_b(1+f)}\mu_b\\
\nonumber\\
\mu_{eff}^{TM}(\omega)&=&(1-f)\mu_{b}+(1-f)\mu^{TM}_{}{b}\\
\nonumber\\
\varepsilon_{eff}^{TM}(\omega)&=&\frac{\varepsilon_a^{TM}(\omega)(1+f)+\varepsilon_b(1-f)}{\varepsilon_a^{TM}(\omega)(1-f)+\mu_b(1+f)}\mu_b
\end{eqnarray}
\endnumparts

\par
\begin{figure}
\centering
\includegraphics{ 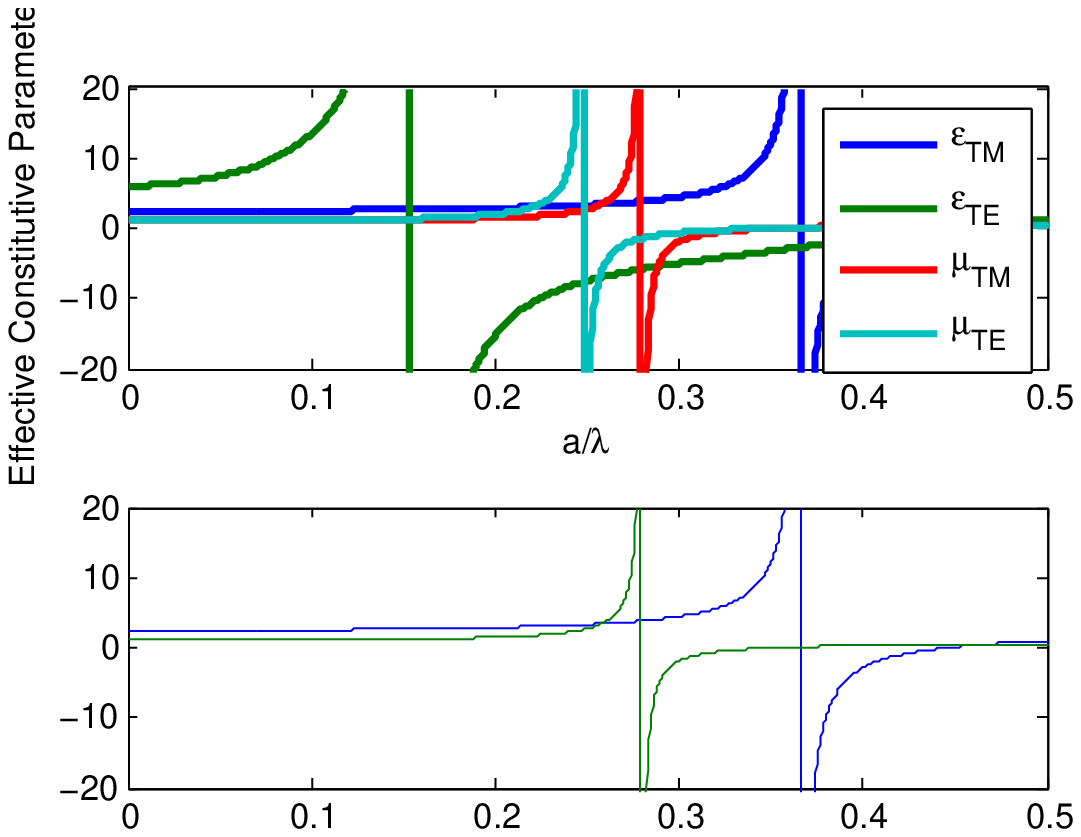}
\caption{\label{fig:emuTETM} Effective constitutive parameters, relative to those of the background, for a square lattice of dielectric cylinders such that $R_a=0.4a$ and $\varepsilon_a=11\varepsilon_b$. The resonances of the permeability and permittivity for the TM case are so sharp that they do not allow to the system to present negative
refraction for that polarization. However the TE polarization is allowed because of the wide range of negative permittivity.}
\end{figure}

 
Figure \ref{fig:emuTETM} depicts a plot of the effective constitutive parameters for a square lattice of dielectric cylinders with $\varepsilon_a=11\varepsilon_b$ and $R_a=0.4a$. Note
that although $\mu_a=\mu_b=\mu_0$ there is a strong magnetic resonance for both polarizations. However we see that the resonance of $\varepsilon^{TM}$ is beyond the homogenization limit 
($\lambda \lesssim 4a$), so that probably this resonance could not be observed.
\par 
\begin{figure}
\centering
\includegraphics{ 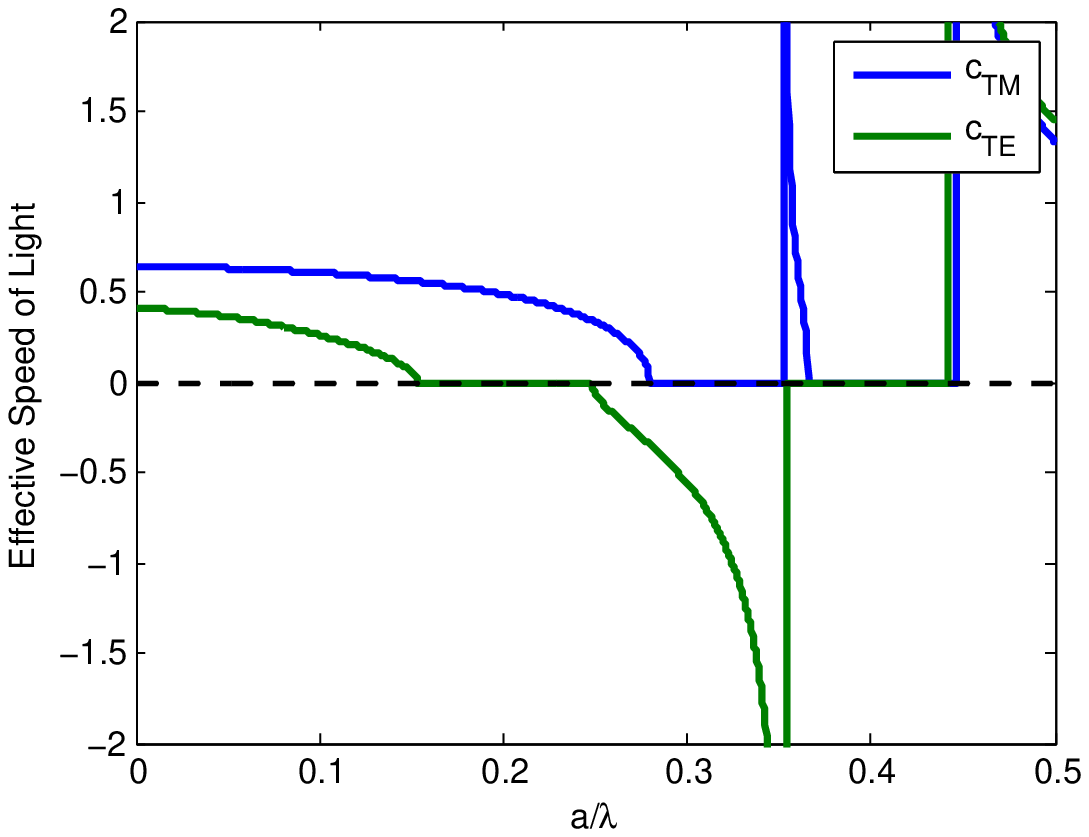}
\caption{\label{fig:ceffTETM}Effective speed of light normalized to that of the background for both TE and TM modes. As we can see, we cannot have negative refraction for the TM mode (see text). The regions
where there is not solution for the speed of light corresponds to those in which $\varepsilon$ and $\mu$ have different sign, so that the speed of sound is imaginary.}
\end{figure}

Figure \ref{fig:ceffTETM} shows the effective speed of light (relative to that of the background) for this system.Note that only the TE polarization presents negative speed of light (or negative refraction index). This
phenomenon is due to the fact that the resonance of $\varepsilon^{TM}$ is too far and too sharp to interact with that of $\mu^{TM}$. If we still want to have negative refraction in the two polarizations we can use anisotropic cylinders, as we did for the acoustic case. 
\par 
If the cylinder's permittivity is given by a tensor of the form 
\begin{equation}
\hat{\varepsilon}_a=(\varepsilon_{ar},\varepsilon_{a\theta},\varepsilon_{az})
\end{equation}
the expressions for the frequency-dependent constitutive parameters are now
\numparts
\begin{eqnarray}
\frac{\varepsilon_{b}}{\varepsilon_{a}^{TE}(\omega)}=\frac{k_b^2R_a^2}{2}\ln k_bR_a+\frac{k_{a}^{TE}R_{a}}{2}\frac{J_{0}(k_{a}^{TE}R_{a})}{J_{1}(k_{a}^{TE}R_{a})}\frac{\varepsilon_{b}}{\varepsilon_{az}}\\
\nonumber\\
\frac{\mu_{a}^{TE}(\omega)}{\mu_{b}}=\frac{1}{k_{a}^{TE}R_{a}}\frac{J_{1}(k_{a}^{TE}R_{a})}{J'_{1}(k_{a}^{TE}R_{a})}\frac{\mu_{a}}{\mu_{b}}\\
\frac{\mu_{b}}{\mu_{a}^{TM}(\omega)}=\frac{k_b^2R_a^2}{2}\ln k_bR_a+\frac{k_{a}^{TM}R_{a}}{2}\frac{J_{0}(k_{a}^{TM}R_{a})}{J_{1}(k_{a}^{TM}R_{a})}\frac{\mu_{b}}{\mu_{a}}\\
\nonumber\\
\frac{\varepsilon_{a}^{TM}(\omega)}{\varepsilon_{b}}=\frac{1}{k_{a}^{TM}R_{a}}\frac{J_{\gamma}(k_{a}^{TM}R_{a})}{J'_{\gamma}(k_{a}^{TM}R_{a})}\frac{\varepsilon_{a\theta}}{\varepsilon_{b}}
\end{eqnarray}
\endnumparts
where $k_{a}^{TE}=\omega\sqrt{\varepsilon_{za}\mu_a}$, $k_{a}^{TM}=\omega\sqrt{\varepsilon_{a\theta}\mu_a}$ and $\gamma^2=\varepsilon_{a\theta}/\varepsilon_{ar}$.
\par 
These expressions allow us to shift the resonance of $\varepsilon^{TM}$ to lower values just by increasing the $\varepsilon_{ar}$, keeping the rest of the system unaltered, in the 
same way as we did for the acoustic case. Note that the equivalence is $\rho_r\to\varepsilon^{TM}_\theta$ and $\rho_\theta\to\varepsilon^{TM}_r$, that is we need to increase the 
value of $\varepsilon_{ar}$ in the electromagnetic case. Thus, as we increase it, the anisotropy factor $\gamma$ goes to 0, and the resonance moves to the left. 
\begin{figure}
\centering
\includegraphics{ 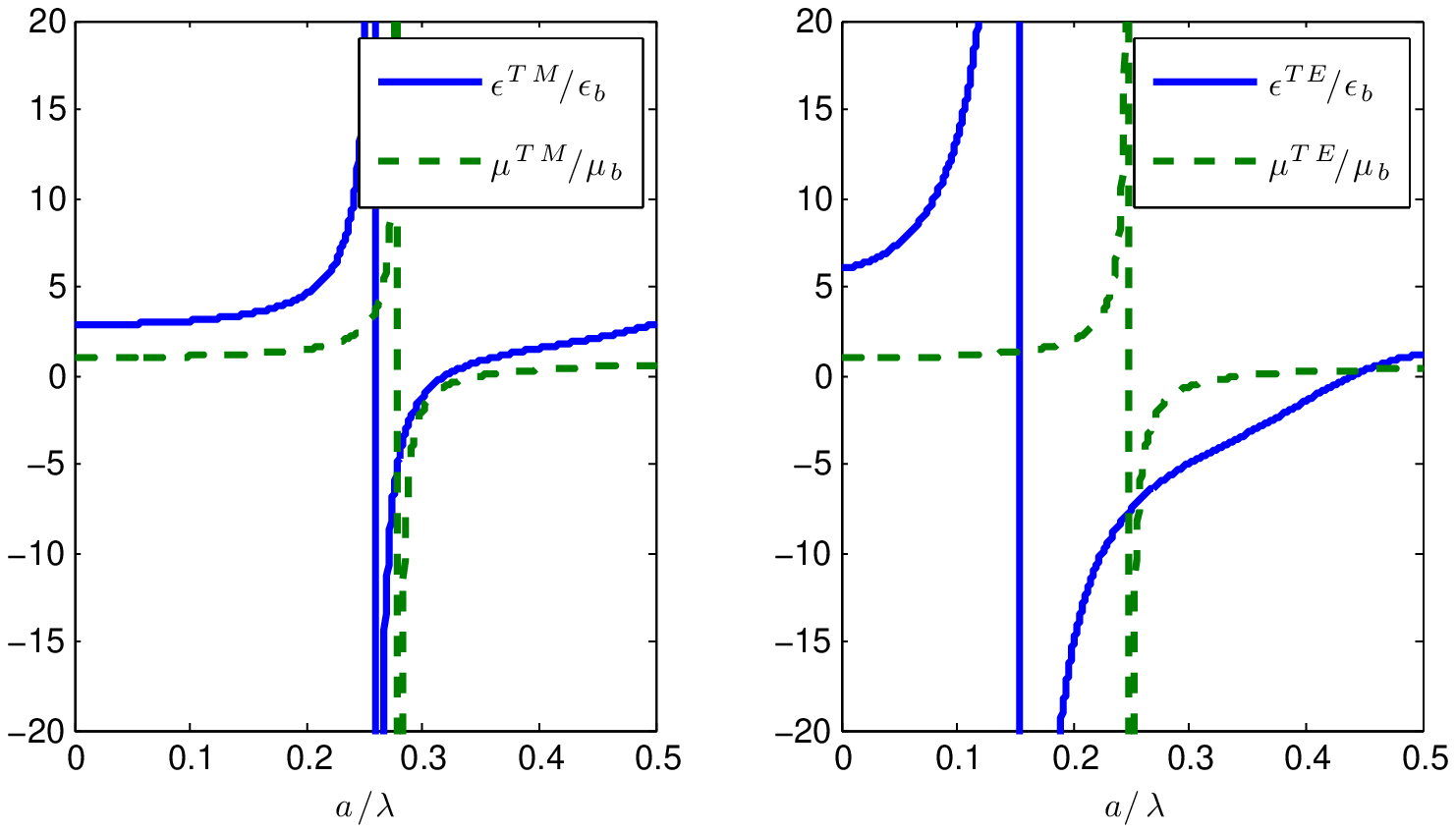}
\caption{\label{fig:aniemuTETM} Frequency shift to the left of $\varepsilon^{TM}$ due to anisotropy in the cylinder. As we see, there is a frequency region in which
all the constitutive parameters for the two modes (TE and TM) are negative, so that we have a region of total negative refraction.}
\end{figure}
\begin{figure}
\centering
\includegraphics{ 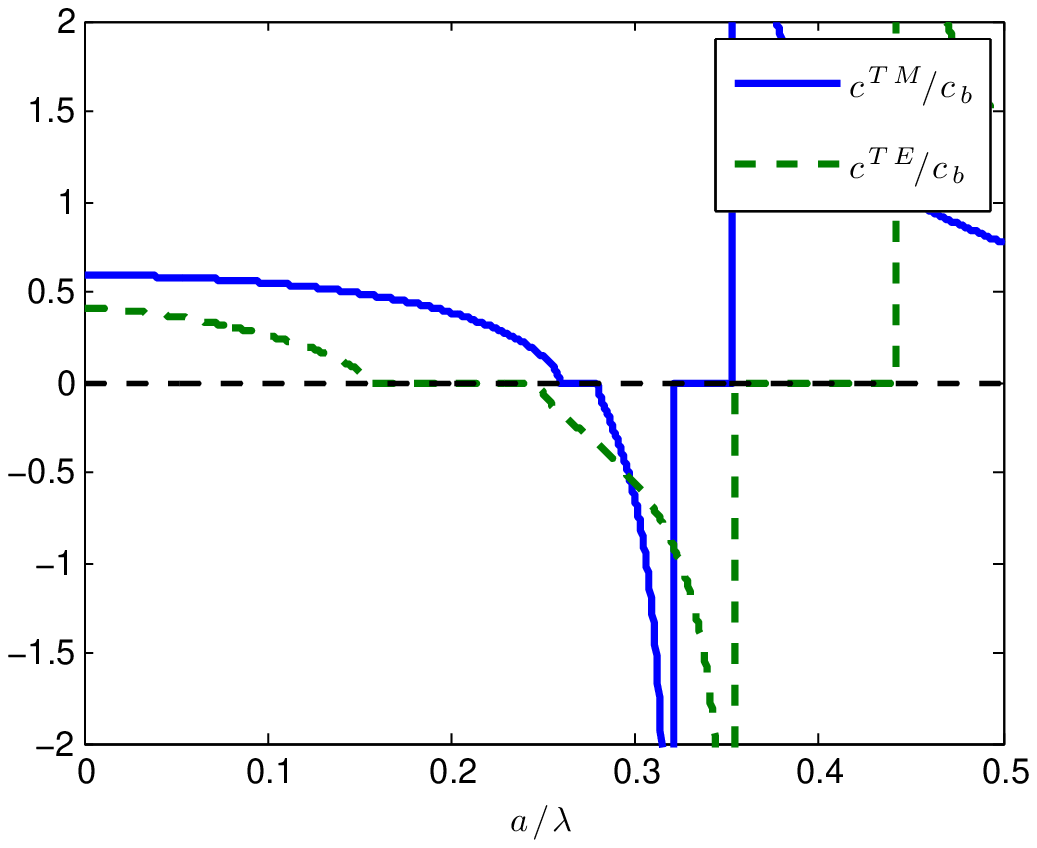}
\caption{\label{fig:aniceffTETM}Effective speed of light for both TE and TM modes for the system of Fig. \ref{fig:aniceffTETM}. It is clear that there is a region of total negative
refraction, thanks to the frequency shift of $\varepsilon^{TM}$ due to anisotropy (see text).}
\end{figure}
\par 
Figure \ref{fig:aniemuTETM} consideres the same system as in figure \ref{fig:emuTETM} but with anisotropic cylinders, where $\varepsilon_r=200\varepsilon_\theta$ and $\varepsilon_\theta=\varepsilon_z=11\varepsilon_b$. Note how all the resonances keep their positions but $\epsilon^{TM}$, which now moves to the left reaching $\mu^{TM}$, leading
therefore to an effective medium with negative refraction. The effective speed of light is depicted in Fig \ref{fig:aniceffTETM}, where it is obvious now that both polarizations
present negative refraction properties within a similar frequency range.
\par 
\section{Summary}
\label{sec:summary}
In summary, a comprehensive multiple scattering formulation of acoustic metamaterials has been introduced. This formulation is based
on a homogenization theory in the quasi-static limit, in which we allow the wave number in the background be arbitrarily small while the wave
number in the scatterers remains finite.
\par
In general, it is shown that ordered or disordered arrays of sound scatterers can behave, in the low frequency limit, like effective fluid-like materials with
either positive or negative acoustic parameters, where these acoustic parameters are the scalar bulk modulus and the tensorial mass density. The behavior of the effective medium depends, among other properties, on the surface fields in the scatterer. Therefore, it
is possible to improve or manage the frequency response of the effective medium with complex scatterers, like fluid-like shells or anisotropic
fluid-like materials. \par
Examples of these scatterers have been analyzed, showing that they present negative effective parameters whenever the theory
predicts them, verifying therefore the formulation presented. Also, it has been shown how these complex scatterers can be used to tune the effective
parameters of the medium.
\par
The theory developed for acoustic waves has been extended to electromagnetic waves in 2D, showing that equivalent type of scatterers can also
be used in electromagnetism to tune the effective medium response. 
\par
In conclusion, the theory presented not only explains the metamaterial behavior found so far in the literature, but also gives the basis for improve this
behavior with more complex scatterers. 
\section*{Acknowledges} 
This work was partially supported by U.S. Office of Naval Research under grant No. N000140910554 and the Spanish Ministry of Science and Innovation under Contracts No. TEC2010-19751 and No. CSD2008-66 (CONSOLIDER Program). Daniel Torrent also acknowledges the contract provided by the program Campus de Excelencia Internacional 2010 UPV.
\appendix
\section{Examples of Scatterers with Local Negative Parameters}
\label{sec:scatterers}
\subsection{Homogeneous and Isotropic Scatterers}
\begin{figure}
\centering
\includegraphics{ 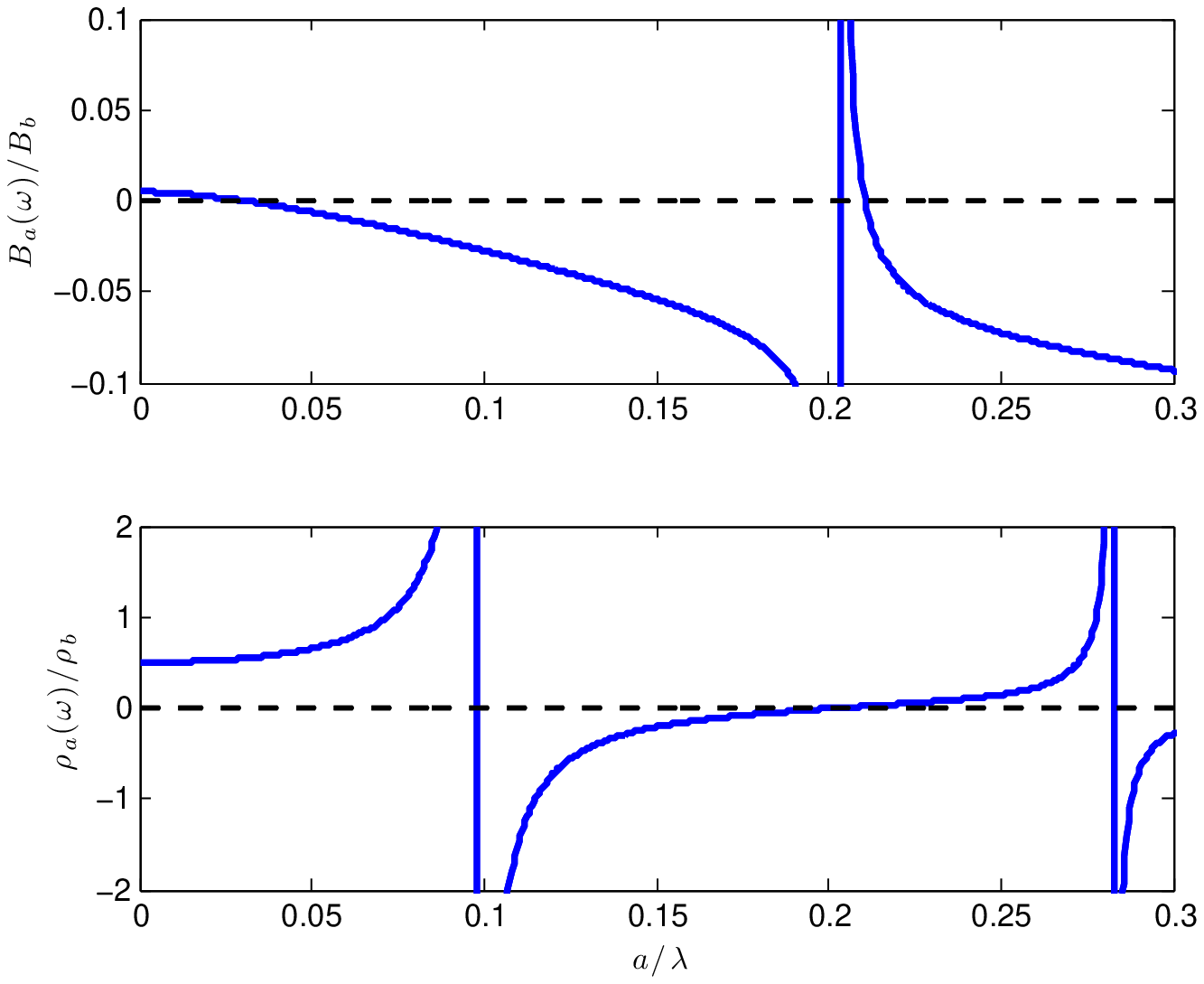}
\caption{\label{fig:rhoBa}Frequency-dependent acoustic parameters of a homogeneous cylinder with $B_{a}=0.005B_{b}$ and $\rho_{a}=0.5\rho_{b}$. The cylinder's radius is $R_{a}=0.3a$, where $a$ is the lattice constant. }
\end{figure}
For a homogeneous scatterer with parameters $\rho_{a}$ and $B_{a}$ the field inside the scatterer is given in terms of Bessel functions. Therefore, after some algebra, the frequency dependent parameters are
\numparts
\begin{eqnarray}
\label{eq:bomegacyl}
B_{a}(\omega)/B_{b}=\frac{k_b^2R_a^2}{2}\ln k_bR_a+\frac{k_{a}R_{a}}{2}\frac{J_{0}(k_{a}R_{a})}{J_{1}(k_{a}R_{a})}\frac{B_{a}}{B_{b}},\\
\label{eq:rhoomegacyl}
\rho_{a}(\omega)/\rho_{b}=\frac{1}{k_{a}R_{a}}\frac{J_{1}(k_{a}R_{a})}{J'_{1}(k_{a}R_{a})}\frac{\rho_{a}}{\rho_{b}},
\end{eqnarray}
\endnumparts
where $k_{a}=\omega\sqrt{\rho_{a}/B_{a}}$. For $k_{a}\to 0$ we recover the cylinder's parameters. Note that the metamaterial
concept appears when $k_{b}<<1$ but $k_{a}$ is not small enough. This condition occurs when $c_{a}<<c_{b}$; i.e. when the wavelength in the background is several times larger 
than that inside the scatterer. 
\par 
Figure \ref{fig:rhoBa} plots the frequency-dependent parameters described by Eqs. \eref{eq:rhoabaomega1} and \eref{eq:rhoabaomega2} for a soft scatterer with $B_{a}=0.005B_{b}$ and $\rho_{a}=0.5\rho_{b}$. These values give a sound speed of the scatterer $c_a=0.1c_b$, which locates the resonances in the low frequency
limit, as can be seen in the figure. 
\par
From \eref{eq:rhoabaomega2} it is deduced that the region of negativity for $\rho_{a}(\omega)$ is determined by the first zeros of 
$J_{1}(k_{a}R_{a})$ and $J_{1}'(k_{a}R_{a})$, which are $\alpha= 3.8317$ and $\alpha'= 1.8412$, respectively\cite{abramowitz1964handbook}, and corresponds to frequencies $\omega_{-}$
such that
\begin{equation}
\frac{1.8412c_{a}}{R_{a}}<\omega_{-}<\frac{3.8317c_{a}}{R_{a}}\quad,\quad \Delta\omega_{-}\approx\frac{2c_{a}}{R_{a}}
\end{equation}
Therefore, if we want to locate the frequency region in the low frequency limit, we have to decrease $c_{a}$ (for fixed $R_{a}$). But, as a consequence, the bandwidth will decrease; that is, the resonance becomes sharper. 
\par
These type of scatterers are possible only in a dense background, like water, where we can get such a low bulk modulus and density. If we need to get metamaterials in an air or gas background another approach should be used instead. 
\subsection{Homogeneous and Anisotropic Scatterers}
Sometimes it is not possible to obtain sound speed smaller than a certain value but we still want to decrease the frequency at which negative mass density appears. In this case,
fluid-like cylinders with circular anisotropy \cite{torrent2010anisotropic} can be used to shift the resonance to lower frequencies.
\par
Anisotropic cylinders are characterized by a scalar bulk modulus $B_a$ and a tensorial mass density whose 
componentes are constant when referred to a cylindrical coordinate system, $\hat{\rho}_a=(\rho_r,\rho_\theta)$.
In these cylinders the pressure field is described in terms of Bessel functions of real order $\gamma q$, where $\gamma=\sqrt{\rho_r/\rho_\theta}$ is the anisotropy factor,
\begin{equation}
\psi_{q}(r,\omega)= J_{\gamma q}(k_{a}r)
\end{equation}
\par
Despite being anisotropic cylinders, the circular symmetry of these scatterers make them suitable for applying the theory developed in this work. Thus, when $q=0$ the field distribution is the same than that of an isotropic cylinder (because $J_{\gamma q}$ for $q=0$ is $J_0$), therefore the frequency dependent bulk modulus will be also given by \eref{eq:bomegacyl}. However \eref{eq:rhoomegacyl} now becomes
\begin{equation}
\rho_{a}(\omega)/\rho_{b}=\frac{1}{k_{a}R_{a}}\frac{J_\gamma(k_{a}R_{a})}{J'_\gamma(k_{a}R_{a})}\frac{\rho_{a}}{\rho_{b}}
\end{equation}
and, as we let the anisotropy factor $\gamma$ be smaller, the dipolar resonance $q=1$ becomes closer to the monopolar one, 
\begin{equation}
\lim_{\gamma\to 0}J_{\gamma}(k_{a}r) \approx J_{0}(k_{a}r)
\end{equation}
decreasing therefore the resonant frequency.
\par
\begin{figure}
\centering
\includegraphics{ 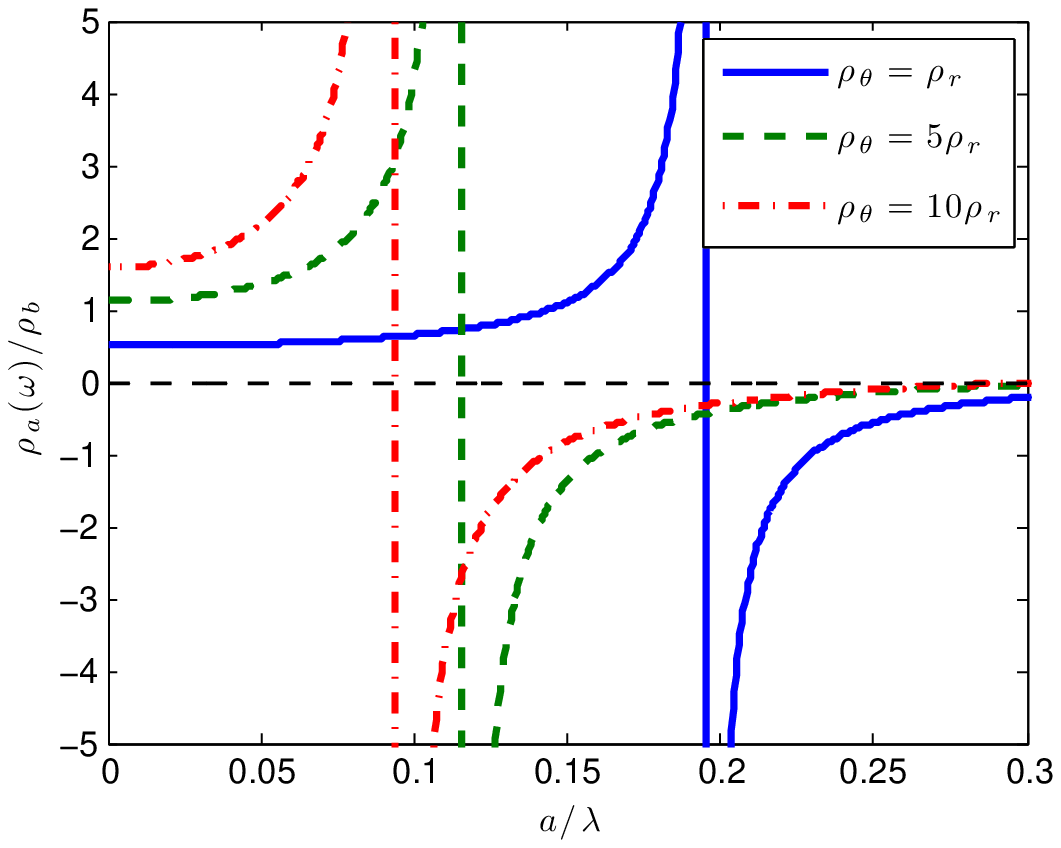}
\caption{\label{fig:rhoa_anisotropic}Frequency-dependent dynamical mass density of an anisotropic fluid like cylinder for three different anisotropy ratios. The radial
component of the sound speed tensor is $c_{r}=0.2c_{b}$ in the three examples. We see that, as we increase the anisotropy ratio, the resonance
is shifted to lower frequencies.}
\end{figure}
In Fig. \ref{fig:rhoa_anisotropic} the frequency-dependent mass density is plotted for three different $\rho_r/\rho_\theta$ ratios. Note how as we increase the value of $\rho_\theta$ (so that we decrease the anisotropy factor $\gamma$) the resonance of the density moves down in frequency.
\par
These type of strongly anisotropic fluid-like cylinders have already been used by Li et al. \cite{li2009hyperlens} for building acoustic hyperlens and it has been recently characterized for different anisotropy ratios in \cite{torrent_tobe}, showing also the same frequency shift.
\subsection{Fluid-Like Shells as Helmholtz Resonators}
Let us assume now that we have a fluid like cylinder of radius $R_a$ and parameters $\rho_a$ and $B_a$. If this cylinder is enclosed by another of radius $R_b>R_a$ and parameters $\rho_s$ 
and $B_s$ we have a fluid-like shell. Obviously this is an idealization, because such a structure cannot be realized with common fluids. 
However, on the one hand, if the shell is an elastic material, this can be sometimes a good approximation and, on the other hand, if the fluids are ``metafluids''
\cite{torrent2007acoustic,pendry2008acoustic,norris2009acoustic}, the structure can be easily fabricated.
\par 
These structures can work, as Helmholz resonators \cite{hu2005two,hu2008homogenization}. Here we give a more rigorous derivation of the resonance frequency.
\begin{figure}
\centering
\includegraphics{ 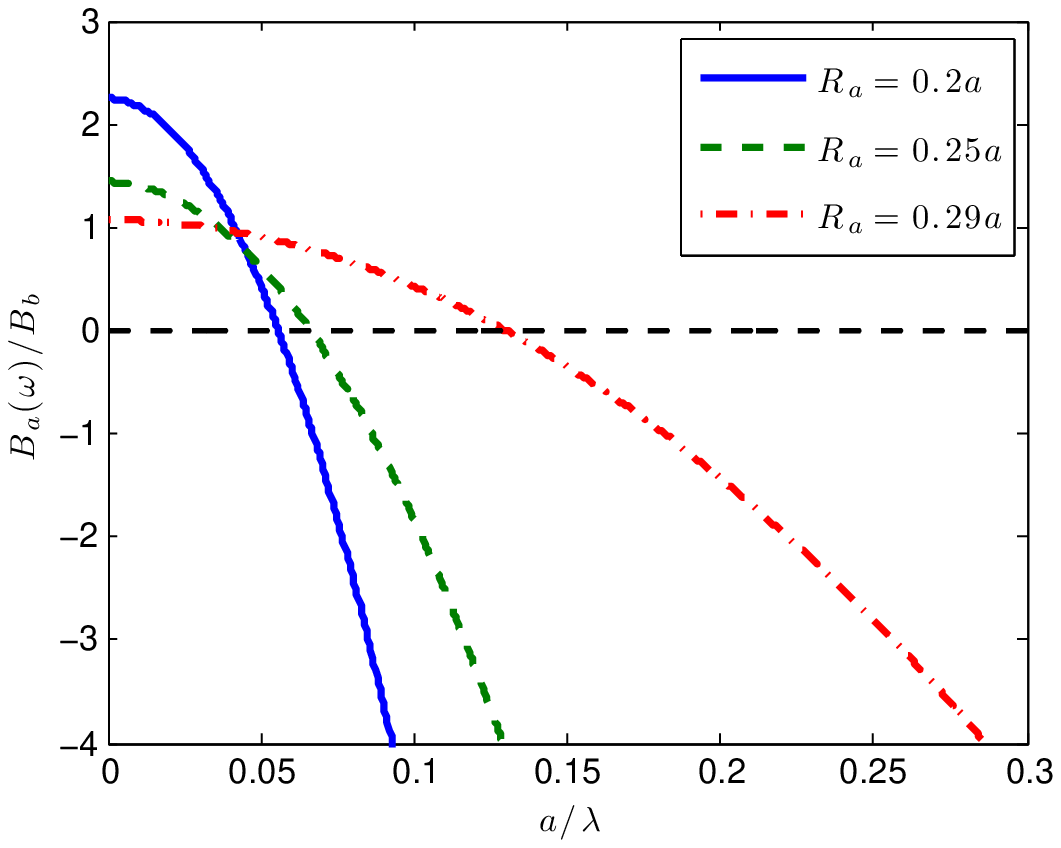}
\caption{Frequency-dependent bulk modulus of a water shell in air with $\rho_s=1000\rho_b$ and $c_{s}=4.4c_{b}$. The outer radius of the
shell is $R_{b}=0.3a$, and the plot shows three different radii $R_{a}$.}
\end{figure}
\par 
After applying the correct boundary conditions, the impedance factor $\chi_q$ of a fluid-like shell is given by 
\begin{equation}
 \chi_q=\frac{\rho_sc_s}{\rho_bc_b}\frac{J_q(k_sR_b)+T_q^aY_q(k_sR_b)}{J'_q(k_sR_b)+T_q^aY'_q(k_sR_b)},
\end{equation} 
where
\begin{equation}
 T_q^a=-\frac{\chi_q^aJ_q'(k_sR_a)-J_q(k_sR_a)}{\chi_q^aH_q'(k_sR_a)-H_q(k_sR_a)},
\end{equation} 
and
\begin{equation}
 \chi_q^a=\frac{\rho_ac_a}{\rho_sc_s}\frac{J_q(k_aR_a)}{J'_q(k_aR_a)}.
\end{equation} 
The parameters of the shell are $\rho_s$ and $c_s$, while those
of the enclosed cylinder are $\rho_a$ and $c_a$.
\par
For $q=0$ and $k_s\to 0$ the inner $T$ matrix $T^a_0$ is equal to
\begin{equation}
T_0^a\approx\frac{\pi R_a^2k_s^2}{4}\frac{\left[1-\frac{B_s}{B_a}\right]}{\xi_a},
\end{equation} 
where
\begin{equation}
 \xi_a=1+\frac{B_s}{B_a}\frac{k_s^2R_a^2}{2}\ln{k_sR_a}.
\end{equation} 
Thus the impedance factor of the shell can be approximated to
\begin{equation}
\chi_0\approx-\frac{2}{k_bR_b}\frac{B_s}{B_b}\frac{1+\frac{k_s^2R_a^2}{2}\left[1-\frac{B_s}{B_a}\right]\ln{k_sR_b}/\xi_a}{1-\frac{R_a^2}{R_b^2}\left[1-\frac{B_s}{B_a}\right]/\xi_a},
\end{equation} 
which can be zero only for $B_s/B_a>>1$  and consequently
\begin{equation}
\xi_a=\frac{k_s^2R_a^2}{2}\frac{B_s}{B_a}\ln{k_sR_b}
\end{equation} 
which defines a cut off frequency $\omega_c$ 
\begin{equation}
\omega_c=\frac{c_a}{R_a}\sqrt{\frac{2\rho_a}{\rho_s\ln(R_b/R_a)}}
\end{equation} 
\par
If the shell is soft, that is, if $B_s/B_a<<1$, a negative bulk modulus appears as in the homogeneous cylinder case,
and the shell nature of the scatterer is not relevant. In that case, as $R_b\to R_a$ the impedance factor reduces to
\begin{equation}
\chi_0\approx-\frac{2}{k_bR_b}\frac{B_s}{B_b}\frac{1}{1-\frac{R_a^2}{R_b^2}\left[1-\frac{B_s}{B_a}\right]}\approx  -\frac{2}{k_bR_b}\frac{B_a}{B_b}
\end{equation}
and the bulk modulus cannot be negative, however we will see now that, in that case, the density can be negative.
\par
When $q=1$ the inner $T$ matrix $T_1^a$ is approximated by
\begin{equation}
T_1^a\approx -\frac{\pi R_a^2k_s^2}{4}\frac{\rho_a-\rho_s}{\rho_a+\rho_s}
\end{equation}
the density becomes negative once the denominator of $\chi_1$ cancels, that is, when 
\begin{equation}
J'_1(k_sR_b)+T_1^aY'_1(k_sR_b)\approx \frac{1}{2}\left[1-\frac{k_s^2R_b^2}{4}-\frac{R_a^2}{R_b^2}\frac{\rho_a-\rho_s}{\rho_a+\rho_s}\right]=0.
\end{equation}
This expression gives a cut off frequency for the negative density of
\begin{equation}
\omega_c^2=\frac{4c_s^2}{R_b^2}\left[1-\frac{R_a^2}{R_b^2}\frac{\rho_a-\rho_s}{\rho_a+\rho_s}\right],
\end{equation}
where now we need that $\rho_s<<\rho_a$.

\section{Technical Details}
\subsection{The anisotropy factor $A$}
A two dimensional periodic array of scatterers is defined by the lattice vectors $\bm{a}_1$ and $\bm{a}_2$, so that the position $\bm{R}_n$ of any scatterer in the lattice can be
determined by two integers $n_1$ and $n_2$ such that
\begin{equation}
\bm{R}_n=n_1\bm{a}_1+n_2\bm{a}_2
\end{equation}
This lattice has also associated the reciprocal lattice vectors $\bm{b}_1$ and $\bm{b}_2$ such that
\begin{equation}
\bm{b}_i\cdot \bm{a}_j=2\pi\delta_{ij}\quad,\quad i=1,2
\end{equation}
If we define the reciprocal lattcie point $\bm{G}_h=(G_h,\theta_h)$ as
\begin{equation}
\bm{G}_h=h_1\bm{b}_1+h_2\bm{b}_2
\end{equation}
the anisotropy factor $A$ can be found in \cite{torrent2008anisotropic} and is given by
\begin{equation}
A=48\sum_{h_1,h_2\neq 0}\frac{J_3(G_hR_{min})}{G_h^3R_{min}^3}e^{-2i\theta_h}
\end{equation}
where $J_3(\cdot)$ is the third order Bessel function and $R_{min}$ is the smaller of $\bm{b}_1,\bm{b}_2,\bm{b}_1+\bm{b}_2$. Factor A can be made always real by properly choosing a coordinate system in which the tensors be diagonal.
\section*{References}

\end{document}